\documentclass[twocolumn,showpacs,preprintnumbers,amsmath,amssymb]{revtex4}

\usepackage{graphicx}
\usepackage{dcolumn}
\usepackage{bm}
\usepackage{natbib}

\begin{document}


\title{Magnetic fluctuation power near proton temperature\\ anisotropy instability thresholds in the solar wind}


\author{S. D. Bale}
\altaffiliation{Physics Department and Space Sciences Laboratory,
University of California, Berkeley}

\author{J. C. Kasper}
\altaffiliation{Harvard-Smithsonian Center for Astrophysics, Cambridge}%

\author{G. G. Howes}
\altaffiliation{Department of Physics and Astronomy,
University of Iowa, Iowa City}

\author{E. Quataert}
\altaffiliation{Physics Department and Astronomy Department,
University of California, Berkeley}

\author{C. Salem}
\altaffiliation{Space Sciences Laboratory,
University of California, Berkeley}

\author{D. Sundkvist}
\altaffiliation{Space Sciences Laboratory,
University of California, Berkeley}

\date{\today}

\begin{abstract}
The proton temperature anisotropy in the solar wind is known to be constrained
by the theoretical thresholds for pressure anisotropy-driven instabilities. 
Here we use approximately 1 million independent measurements of gyroscale magnetic fluctuations in the solar
wind to show for the first time that these fluctuations are enhanced along the temperature anisotropy thresholds of the
mirror, proton oblique firehose, and ion cyclotron instabilities.  In addition, the measured magnetic compressibility
is enhanced at high plasma beta ($\beta_\parallel \gtrsim 1$) along the mirror instability threshold but small elsewhere, 
consistent with expectations of the mirror mode.  The power in this frequency (the 'dissipation') range is often considered to be driven by
the solar wind turbulent cascade, an interpretation which should be qualified in light of the present results.  In particular, we show
that the short wavelength magnetic fluctuation power is a strong function of collisionality, which relaxes the temperature anisotropy away
from the instability conditions and reduces correspondingly the fluctuation power.
\end{abstract}

\maketitle

The physical processes that regulate the expansion of the super-Alfv\'enic solar wind into space
include adiabatic particle motion, plasma instabilities, and binary particle collisions.  As the wind
expands, plasma density $n_p$ and magnetic field $|B|$ decrease radially.  The Chew-Goldberger-Low
(CGL) relations \cite{CHEW:1956p4064} predict that the plasma ions should become anisotropic in 
the sense of $T_\parallel > T_\perp$ if the particle motion is adiabatic and collisionless; here $T$ is the
ion temperature parallel and perpendicular to the background magnetic field.  However, Coulomb collisions
and pressure-anisotropy instabilities act to pitch-angle scatter the plasma back towards isotropy \cite{EVIATAR:1970p199}.  At 1 AU, the most probable
value of the proton temperature anisotropy is $T_\perp/T_\parallel \approx 0.89$ (see Figure 1, top panel below).  If CGL were valid, this would imply
a proton temperature anisotropy of $T_\perp/T_\parallel \ge$ 200 at 5 solar radii.

The same pressure-anisotropy instabilities that operate in the solar wind are believed to operate in other low-collisionality astrophysical plasmas, including clusters of galaxies \citep{Schekochihin:2005p7666} and some accretion disks onto black holes \citep{Sharma:2006p6378, Sharma:2007p7617}.   In the latter environment, these instabilities not only modify the thermodynamics of the plasma (as in the solar wind), but they also play a crucial dynamical role, regulating the anisotropic stress that helps transport angular momentum, allowing accretion to proceed.

Growth of ion temperature anisotropy instabilities has been studied in a (relatively collisional) laboratory device, where isotropization and magnetic
fluctuations were observed corresponding to the Alfv\'en ion cyclotron instability \cite{Scime:2000p5378}.  Similar results were obtained in the
solar wind at 1 AU \citep{Gary:2001p4941} (for $T_\perp/T_\parallel >$ and solar wind speeds greater than 600 km/s), suggesting that the proton cyclotron
instability plays a role.  Both mirror mode and ion cyclotron anisotropy instabilities appear to be active in the terrestrial magnetosheath, inferred from observed
constraints on the temperature anisotropy \citep{ANDERSON:1994p7478, Phan:1996p7473}.

More recently, Kasper et al. \cite{Kasper:2002p858} used measurements by the SWE instrument \cite{Ogilvie:1995p385} on the Wind spacecraft
 to show that the solar wind proton temperature
anisotropy $T_\perp/T_\parallel$ is constrained with respect to proton parallel beta 
$\beta_\parallel (= n_p k_b T_\parallel/B^2$) in a way that is consistent with expectations of the
proton firehose instability (for $T_\perp/T_\parallel < 1$).  Hellinger et al.
(2006) compared linear instability calculations with the measured anisotropies (from this same dataset) and found that the 
constraints on the observed anisotropies were best described by the mirror instability (in contrast to previous results \cite{Gary:2001p4941}) for the case
of perpendicular anisotropy ($T_\perp/T_\parallel > 1$) and the oblique firehose instability for parallel 
anisotropy ($T_\perp/T_\parallel < 1$).  An instability threshold can be derived by calculating marginal stability (in practice
$\gamma = 10^{-3} \omega_{ci}$) for values of $\beta_\parallel$ \cite{Hellinger:2006p299}.

\begin{equation}
\frac{T_\perp}{T_\parallel} = 1 + \frac{a}{(\beta_\parallel - \beta_0)^b}
\end{equation}

Equation (1) generalizes previous results \citep{Gary:1994p7247,Gary:2001p4941} and constrains the Wind/SWE observations for values
of $a, b,$ and $\beta_0$ corresponding to the mirror and oblique firehose instablilities (coefficients are given below).

Here we use experiments on the Wind spacecraft to show
the first direct correspondence of the measured wave power to the anisotropy-driven instabilities.
Wind was launched in 1994 and
has spent long intervals in the solar wind, often near the first Lagrange point upstream of the Earth.
The data set spans the time interval Nov. 21, 1994 to Nov. 12, 2004 and includes 1,026,112 independent measurements of solar wind
 plasma and magnetic field.
Proton density $n_p$, velocity $\vec{v}_{sw}$, and temperature are measured by the 
Faraday cup instrument of SWE (Solar Wind Experiment) \cite{Ogilvie:1995p385}.  Both the
 parallel proton temperature $T_\parallel$ and perpendicular temperature $T_\perp$ are computed
 by comparison with the average magnetic field direction.  The Magnetic Field Investigation (MFI)
 is used to measure the solar wind magnetic field \cite{Lepping:1995p6447} at either 22 vectors/s or 11 vectors/s depending on telemetry mode and averaged to 
 3 second intervals; we denote this 3 second average data as $\vec{B}$.   The vector rms fluctuation field
 during the 3 second average interval is denoted as $\delta \vec{B}$.  Each
 fluctuation field measurement $\delta \vec{B}$ is rotated into a coordinate system defined by the average field direction
 $\hat{B}$, so that we have both the compressive component $\delta B_\parallel$ and the shear component
 $\delta B_\perp(=(\delta B_{\perp,1}^2 + \delta B_{\perp,2}^2)^{1/2}$).  We 
 then define the magnetic compressibility as
 ${\delta B_\parallel^2}/({\delta B_\parallel^2 + \delta B_\perp^2})$ \cite{KRAUSSVARBAN:1994p2293}.  The magnetic field is an rms measurement over the bandwidth $\Delta f = $ (0.3 - 5.5) Hz,  or (0.3 - 11) Hz in high TM mode.  Since the magnetic fluctuation spectrum falls as $f^{-5/3}$ (or steeper) at these frequencies,
 the power in this bandwidth is dominated by the lowest measured frequency $f_0 \approx$ 0.3 Hz.  Since it is believed
 that $k_\perp \gg k_\parallel$ \citep{GOLDREICH:1995p7765}, the natural frequencies of the turbulence are all much lower (i.e. $f_0 \gg f_{ci}$) and Taylor's
 hypothesis applies ($\omega = k v_{sw}$).  Therefore the measured power at $f_0 \approx$ 0.3 Hz corresponds to power at wavenumber
 $k \rho_i \approx (f_0/f_{ci}) (v_{th}/v_{sw})$.  The distribution (in our data) of this parameter is sharply peaked at $k \rho_i \approx$ 0.56 with a half-width
 of 0.32; therefore these measurements correspond to magnetic fluctuation power at $k \rho_i \approx 0.56 \pm 0.32$.
%
%

\begin{figure}
\includegraphics[width=90mm,height=180mm, scale=1,clip=true, draft=false]{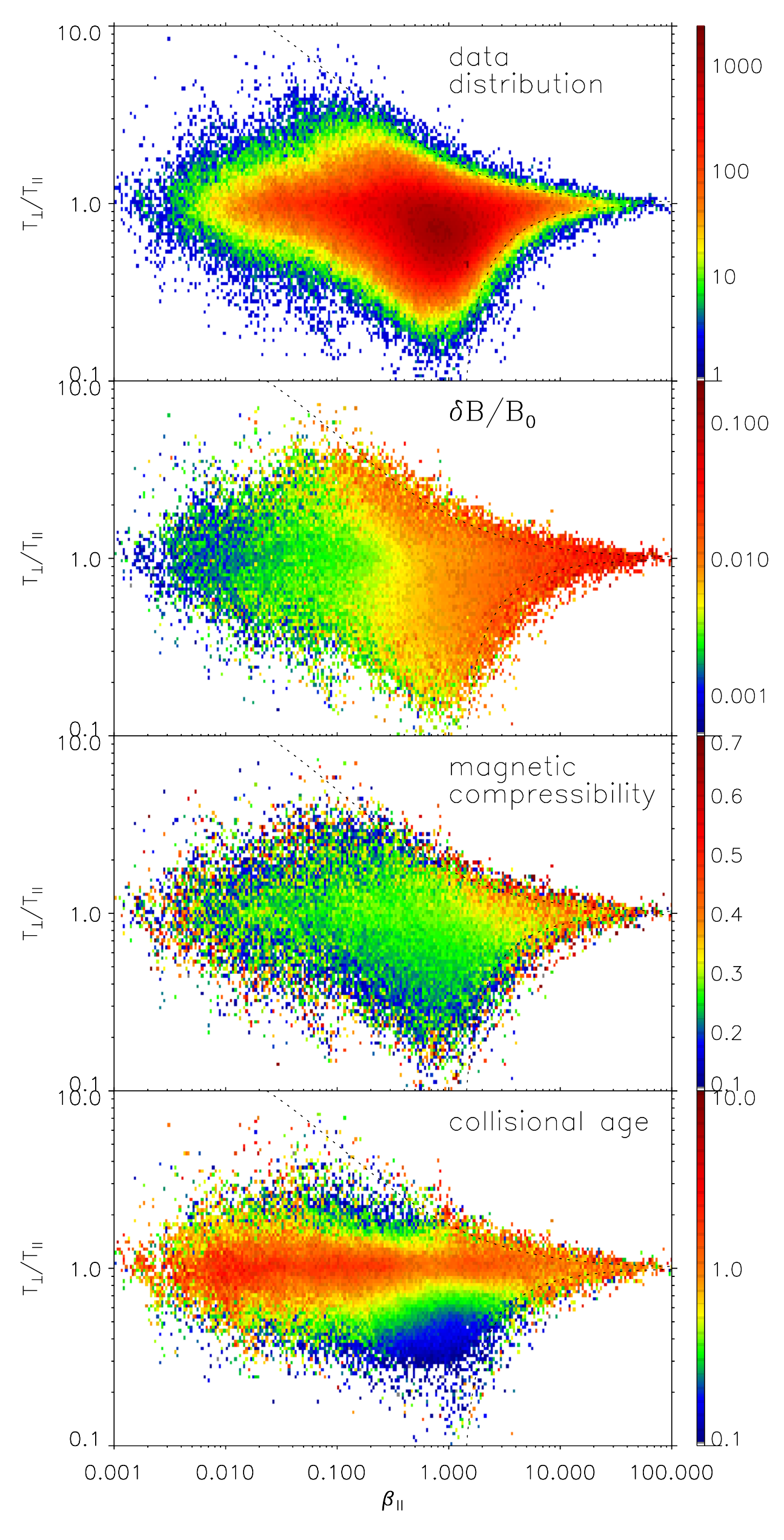}
\caption{The distribution of proton temperature anisotropy ($T_\perp/T_\parallel$) measurements with respect to the 
parallel plasma beta $\beta_\parallel$ (upper panel) is constrained by the oblique proton firehose instability
threshold (lower dotted line) and the mirror instability threshold (upper dotted line).  In the second panel,
the magnetic fluctuation amplitude $|\delta B|/|B|$ is shown to be enhanced along the instability thresholds
and overall at high $\beta_\parallel$ where the thresholds converge.  The third panel shows magnetic compressibility
$\delta B_\parallel^2/(\delta B_\parallel^2+\delta B_\perp^2)$, which is enhanced at high $\beta_\parallel$ ($>$1) along the
mirror threshold, as expected for the mirror instability.  The lower panel shows the 'collisional age', which is largest
around $T_\perp/T_\parallel \approx$ 1 suggesting that isotropy results largely from Coulomb collisions.  Anisotropic plasma
is relatively collisionless.}
\end{figure}

The upper panel of Figure 1 shows the distribution of measurements of proton temperature anisotropy
($T_\perp/T_\parallel$) against parallel proton plasma beta $\beta_\parallel$; this distribution shows the striking
signatures of the regulation of the anisotropy by instabilities.  These proton measurements are a subset of 
those used by Kasper et al.\cite{Kasper:2002p858}; here we include only proton measurements at times when there also exist good
measurements of $\delta \vec{B}$.  Dotted lines (on all four panels) show the instability thresholds for the mirror instability at $T_\perp/T_\parallel > 1$
(equation 1 with $(a, b, \beta_0) = (0.77, 0.76, -0.016)$) and the oblique firehose instability at $T_\perp/T_\parallel < 1 $(equation 1 with $(a, b, \beta_0) = (-1.4, 1.0, -0.11)$) \cite{Hellinger:2006p299}.
The second panel of Figure 1 shows the average measured amplitude of 
the magnetic fluctuations $|\delta \vec{B}|/|\vec{B}|$ in the space of ($\beta_\parallel, T_\perp/T_\parallel$), 
as in the upper panel.  A general trend of enhanced fluctuations with larger $\beta_\parallel$ is clearly visible.
Furthermore, the fluctuation amplitude is enhanced along both the mirror/IC (for $T_\perp/T_\parallel > 1$) and
oblique firehose ($T_\perp/T_\parallel < 1$) thresholds.  This is strong evidence that these instabilities are excited here and act to isotropize the plasma; this is the principal result of this study.  The third panel of Figure 1 shows
the magnetic compressibility ${\delta B_\parallel^2}/({\delta B_\parallel^2 + \delta B_\perp^2})$, which is
enhanced to values of $\sim 0.3$ for compressive solar wind ($T_\perp/T_\parallel > 1$) with $\beta_\parallel \gtrsim 1$,
as would be expected for the mirror instability.  The compressibility becomes smaller for $\beta_\parallel < 1$, which is consistent with the
Alfv\'en ion cyclotron mode, however the power continues to be bounded by the mirror threshold.  Linear mirror instability calculations \cite{KRAUSSVARBAN:1994p2293} for $T_\perp/T_\parallel > 1$ predict
values of the magnetic compressibility between 0.8 and 1, therefore our measurements 
suggest a mixture of waves.  Furthermore, it is interesting to note that the typical value of the magnetic compressibility
{\em away} from the thresholds is small $\sim$ 0.1, in contrast to theories which suggest evolution of the MHD cascade to
the (compressive) magnetosonic/whistler branch at short wavelengths \cite{Gary:2008p6467}.  The bottom panel of Figure 1 shows the average collisional 'age'
in each $(\beta_\parallel, T_\perp/T_\parallel)$ bin; the collisional age $\tau$ is defined as $\tau = \nu_{pp} ~L/v_{sw}$, the Coulomb proton-proton collision frequency $\nu_{pp}$
multiplied by the transit time from the Sun to 1 AU and is an estimate of the number of binary collisions in each
plasma parcel during transit from the Sun to the spacecraft.  It is interesting, however obvious, that the more collisional
plasma is more isotropic; away from $T_\perp/T_\parallel \approx 1$ the plasma is relatively collisionless.  It has been shown recently
\citep{Kasper:2008p7149} that collisional age organizes solar wind instabilities better than the traditional distinction of 'fast' and 'slow'
wind.

%

\begin{figure}[h]
\includegraphics[width=92mm,height=68mm, scale=1,clip=true, draft=false]{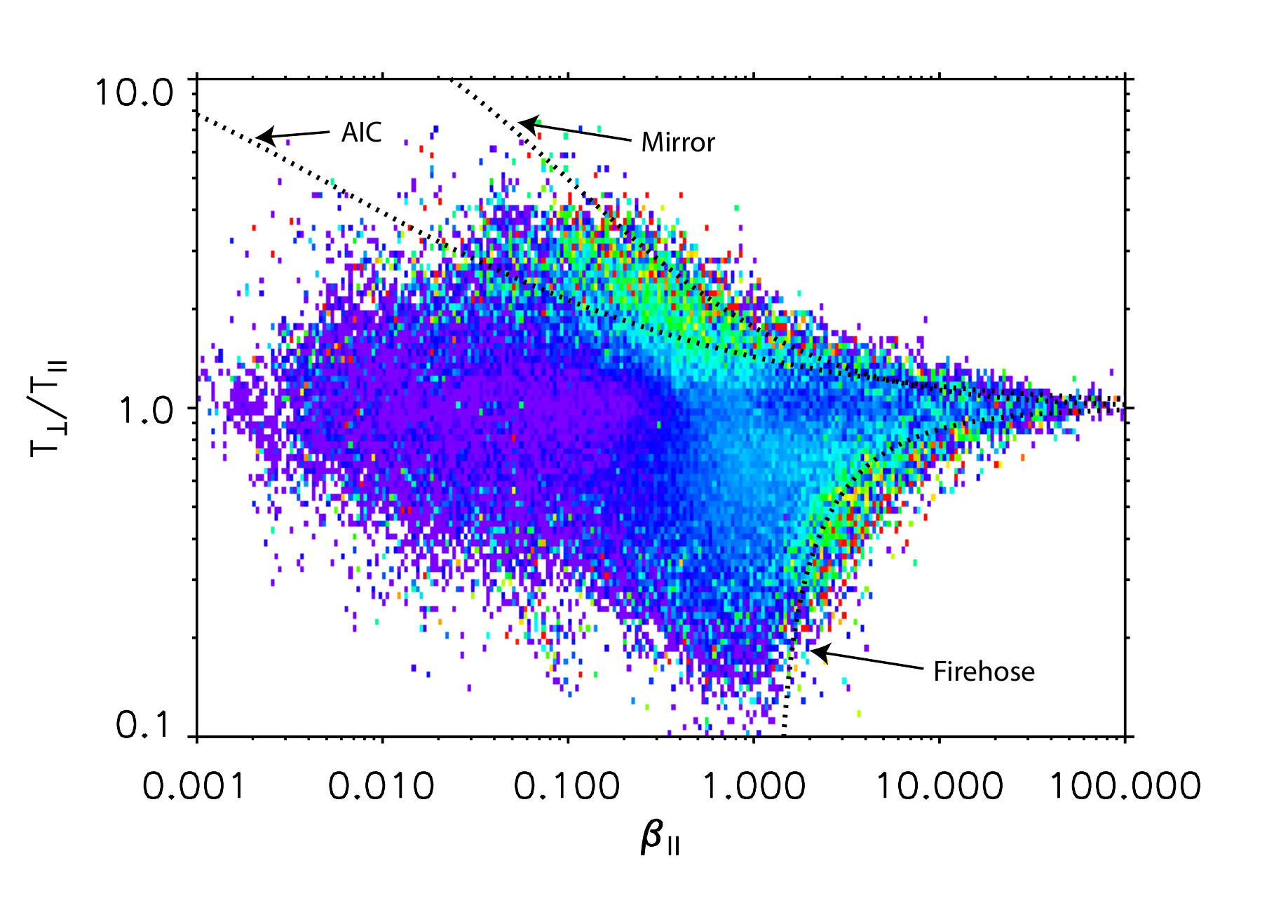}
\caption{The magnitude of magnetic fluctuations $|\delta B|$ averaged into bins of $T_\perp/T_\parallel$ vs $\beta_\parallel$.
Enhanced power exists well away from the thresholds, as expected.  The regions of enhanced $\delta B$ corresponds to the enhanced proton heating in {\it Liu et al} (2006).}
\end{figure}

Figure 2 shows the magnetic fluctuations data $|\delta B|$ unnormalized by $B_0$.  Linear instability thresholds
associate with the mirror, ion cyclotron (AIC) , and oblique firehose instabilities \cite{Hellinger:2006p299} are overlaid.  
  It is interesting to note that, as found by Hellinger et al \cite{Hellinger:2006p299},
the oblique firehose and the mirror instabilities appear to constrain the observed distribution of data, not the ion cyclotron
nor parallel firehose instabilities (in spite of a larger growthrate); this may be because both the mirror and oblique firehose
are non-propagating instabilities.  The regions of enhanced magnetic
fluctuations, near the mirror and oblique firehose thresholds also correspond to measurements of enhanced proton
temperature published elsewhere \cite{Liu:2006p860}.  It is unclear if this indicates plasma heating due to anisotropy instabilities, in 
addition to pitch-angle scattering, or if the 'younger' (less collisional) plasma is merely hotter than average.  

%
%

\begin{figure}[h]
\includegraphics[width=90mm,height=65mm, scale=1,clip=true, draft=false]{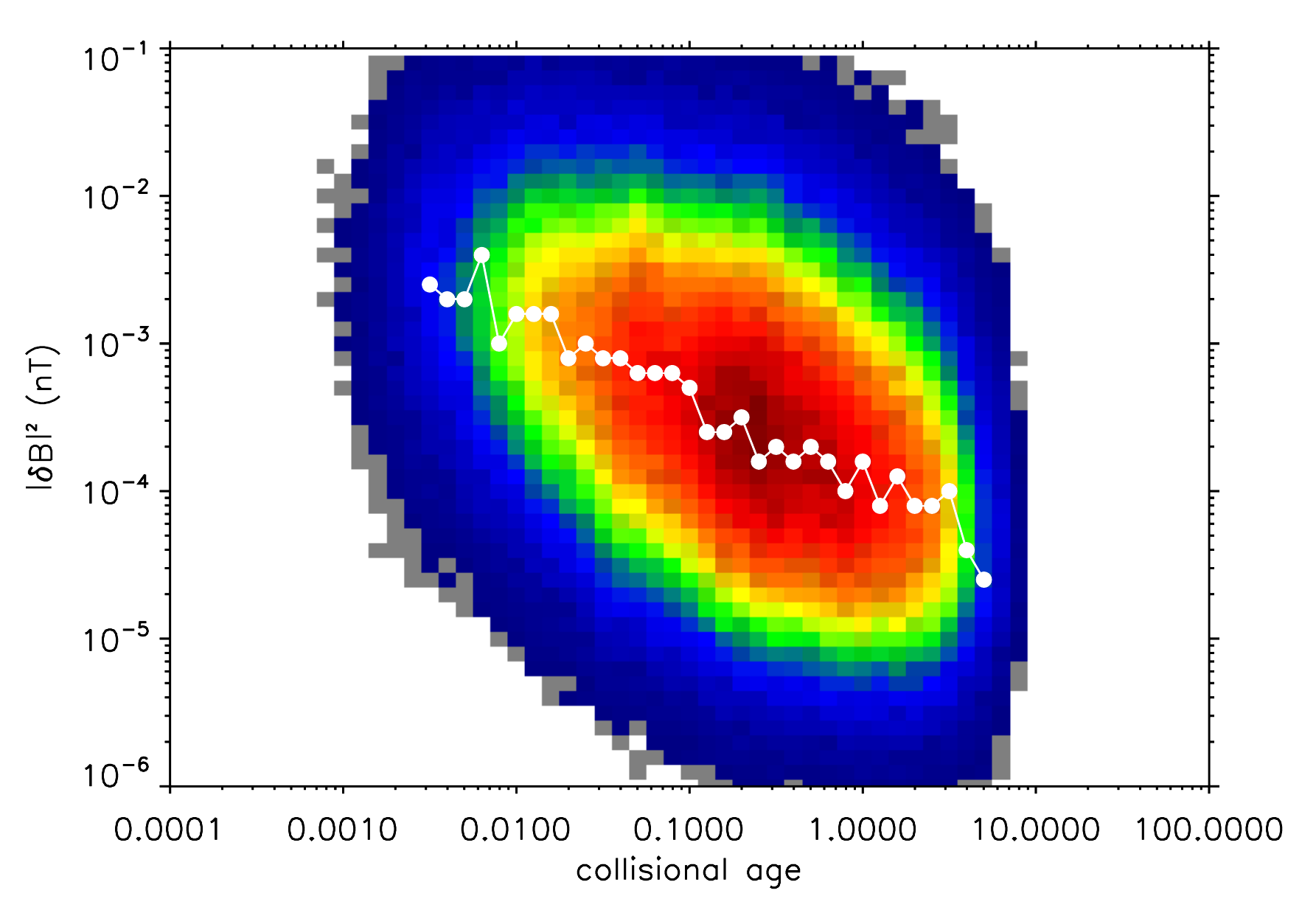}
\caption{Magnetic fluctuation amplitude $|\delta B|^2$ as a function of the collisional age; the white dots are the most likely value
of $|\delta B|^2$ in each age bin.  Magnetic fluctuations near the proton gyroradius ($k \rho \approx 1/2$) are suppressed in more collisional plasma.  Coulomb
collisions maintain the isotropy of the protons, which then remain far from the instability thresholds. Note that this corresponds to a factor of
100 suppression of magnetic power $\delta B^2$ over the full range of collisionality.}
\end{figure}

Figure 3 shows histograms of the fluctuation amplitude squared $|\delta B|^2$ in bins of collisional age; the white dots show the most
probable value.  The overall magnetic fluctuation power $\delta B^2$ is a function of the collisional age, with the magnetic power
 weaker by a factor of $\sim$100 for more collisional plasma.  This effect
is a proxy for the temperature anisotropy:  collisional plasma is more isotropic and, therefore, further from the instability
thresholds.  This underscores the important point that the power spectral density (PSD) of magnetic fluctuation power near 1 Hz in the solar wind is
modified by these local instabilities.  Previous studies of the PSD of short wavelength solar wind turbulence have not accounted for this and should be reexamined \cite{Leamon:1999p5236, Bale:2005p98, Alexandrova:2008p7187, Horbury:2008p4750, Podesta:2009p7052}.  If the $|\delta B|^2$ values in Figure 3 are divided by the measurement bandwidth (approximately 5-10 Hz depending on sample rate), they can be compared to power spectral density (PSD) measurements published previously (over the bandwidth of 0.3 to 11 Hz), noting that the power is dominated by the amplitude at the lowest frequencies (0.3 Hz).  A log-log fit to the most probable
values in Figure 3 gives a relationship $|\delta B|^2 \approx 10^{-4} \tau^{-0.57}$, where $\tau = \nu_{pp} ~L/v_{sw}$ is the collisional age.

Figure 4 shows wavelet power spectral density for 3 distinct time intervals; these spectra are shown plotted against $k \rho_i$ assuming
Taylor's hypothesis ($\omega = k v_{sw}$) and using the local values of $v_{sw}, v_{th}$, and $\omega_{ci}$ and scaling the spectra to have similar power over the
range $k \rho_i \in (0.2, 1.0)$.   The black trace in Figure 4 shows the magnetic power from an interval of isotropic plasma $(T_\perp/T_\parallel, \beta_\parallel) \approx (1, 0.7)$, 
the red trace is data from an interval of perpendicular anisotropy $(T_\perp/T_\parallel, \beta_\parallel) \approx (2.2, 0.2)$, and the blue trace from an interval of parallel anisotropy $(T_\perp/T_\parallel, \beta_\parallel) \approx (0.5, 1.9)$.  Although wavenumbers below $k \rho_i \approx$ 0.2 are aliased by the Morlet wavelet transform and are unreliable.
The inertial range, where the power is proportional to $k^{-5/3}$ (shown by a solid black line in Figure 4), can be made to agree, the nature of
the short wavelength spectrum above $k \rho_i \approx 1$ varies between spectra.  While this result is not statistical, nor is it surprising given the clear organization 
of the magnetic fluctuation power in Figure 1, it begs a reexamination of 'dissipation range' turbulence.
Future observational studies of turbulent dissipation and secondary (kinetic Alfv\'en wave) cascades should attempt to understand the role of these
local instabilities, which may confuse the interpretation of solar wind magnetic power spectra.

%
%
\begin{figure}[]
\includegraphics[width=90mm,height=80mm, scale=1,clip=true, draft=false]{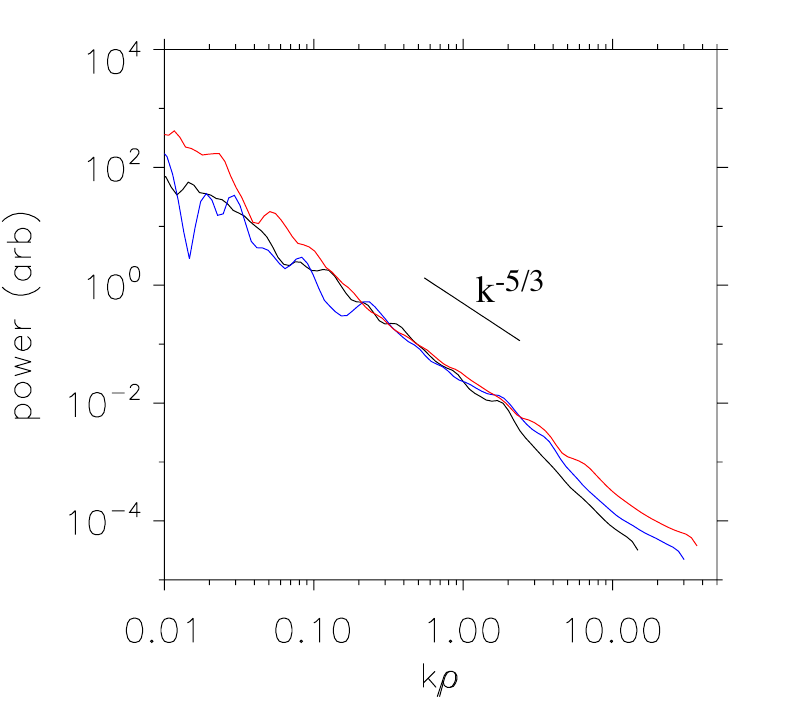}
\caption{Wavelet power spectra of magnetic fluctuations at three different time intervals corresponding to perpendicular
anisotropy $T_\perp/T_\parallel$ = 2.2, $\beta_\parallel$ = 0.2 (red trace), parallel anisotropy $T_\perp/T_\parallel$ = 0.5, $\beta_\parallel$ = 1.9 (blue trace) and isotropic protons $T_\perp/T_\parallel$ = 1, $\beta_\parallel$ = 0.7 (black trace).  The spectra
are plotted against $k \rho$ and have been scaled to the same average value over the interval $k \rho \in (0.2, 1.0)$, as 
described in the text.  A solid black line with scaling $k^{-5/3}$ is shown.  The interval
above $k \rho \approx 1/2$ corresponds to the statistical data shown in Figures 1 and 2.}
\end{figure}

\bibliographystyle{apsrev}


\end{document}